\documentclass[11pt,epsf,letterpaper]{article}%
\usepackage{setspace}
\usepackage{color}
\usepackage{amsmath}
\usepackage{amsfonts}
\usepackage{verbatim}
\usepackage{amssymb}
\usepackage{graphicx}%
\providecommand{\U}[1]{\protect\rule{.1in}{.1in}}
\begin{document}

\title{Gribov pendulum in the Coulomb gauge on curved spaces}
\author{Fabrizio Canfora$^{1}$, Alex Giacomini$^{2}$ and Julio Oliva$^{2}$\\$^{1}$\textit{Centro de Estudios Cient\'{\i}ficos (CECS), Casilla 1469,
Valdivia, Chile.}\\$^{2}$\textit{Instituto de F\'{\i}sica, Facultad de Ciencias, Universidad
Austral de Chile, Valdivia, Chile.}\\{\small canfora@cecs.cl, alexgiacomini@uach.cl, julio.oliva@docentes.uach.cl}}
\maketitle

\begin{abstract}
In this paper the generalization of the Gribov pendulum equation in the
Coulomb gauge for curved spacetimes is analyzed on static spherically
symmetric backgrounds. A rigorous argument for the existence and uniqueness of
solution is provided in the asymptotically AdS case. The analysis of the
strong and weak boundary conditions is equivalent to analyzing an effective
one-dimensional Schr\"{o}dinger equation. Necessary conditions in order for
spherically symmetric backgrounds to admit solutions of the Gribov pendulum
equation representing copies of the vacuum satisfying the strong boundary
conditions are given. It is shown that asymptotically flat backgrounds do not
support solutions of the Gribov pendulum equation of this type, while on
asymptotically AdS backgrounds such ambiguities can appear. Some physical
consequences are discussed.

\end{abstract}

\section{Introduction}

The Yang-Mills Lagrangian $L$ is one of the basic blocks of the Standard
Model:
\begin{equation}
L=trF_{\mu\nu}F^{\mu\nu}\;,\;\;\;(F_{\mu\nu})^{a}=(\partial_{\mu}A_{\mu
}-\partial_{\mu}A_{\mu}+[A_{\mu},A_{\nu}])^{a}\ .
\end{equation}
The degrees of freedom of the theory are encoded in the connection$\ (A_{\mu
})^{a}$, which is a Lie algebra valued one form. The action functional is
invariant under finite gauge transformations, which act on the gauge potential
as%
\begin{equation}
A_{\mu}\rightarrow U^{\dagger}A_{\mu}U+U^{\dagger}\partial_{\mu}U
\label{gaugetranformation}%
\end{equation}
whereas the physical observables are invariant under proper gauge
transformations. The latter has to be everywhere smooth and it has to decrease
fast enough at infinity such that a suitable norm, to be specified later,
converges\footnote{A key reference on the problem of defining a proper gauge
transformation is \cite{Benguria:1976in}.}. This invariance is related with
the existence of first class constraints, which in turn imply that the degrees
of freedom of the theory are less than the number of algebraically independent
components of the gauge potential.

Up to now, the program of using from the very beginning gauge invariant
variables, has been completed only in the cases of topological field theories
in 2+1 dimensions \cite{wittenjones}, while it is still far from clear how to
perform practical computations in a completely gauge-invariant way for
Yang-Mills theories in 2+1 and 3+1 dimensions. Furthermore, the gauge-fixing
problem is also relevant in the classical theory since, when using the Dirac
bracket formalism, the Faddeev-Popov determinant appears in the denominators
of the Dirac-Poisson brackets (see, for instance, the detailed analysis in
\cite{HRT}).

A gauge fixing condition is the common practical solution, the most convenient
choices being the Coulomb gauge and the Lorenz gauge\footnote{Other gauge
fixings are possible such as the axial gauge, the temporal gauge, etc.,
nevertheless these choices have their own problems (see, for instance,
\cite{DeW03}).}:%
\begin{equation}
\partial^{i}A_{i}=0,\;\;\;\partial^{\mu}A_{\mu}=0\ ; \label{coulomblorenz}%
\end{equation}
where $i=1,..,D$ are the spacelike indices and $\mu=0,1,...,D$ are space-time indices.

This procedure has enormous value, allowing perturbative computations around
the trivial vacuum $A_{\mu}=0$. However, the existence of a proper gauge
transformation (\ref{gaugetranformation}) preserving one of the conditions
(\ref{coulomblorenz}) would spoil the whole quantization procedure. In
\cite{Gri78}, Gribov showed that\footnote{Furthermore, it has been shown by
Singer \cite{singer}, that if Gribov ambiguities occur in Coulomb gauge, they
occur in all the gauge fixing conditions involving derivatives of the gauge
field.} a \textit{proper gauge fixing} is not possible.

In the path integral formalism, an ambiguity in the gauge fixing corresponds
to smooth zero modes of the Faddeev-Popov (\textbf{FP}) operator satisfying
suitable boundary conditions. In order to define the path integral in the
presence of Gribov copies, it has been suggested to exclude classical $A_{\mu
}$ backgrounds which generate zero modes of the FP operator (see, in
particular, \cite{Gri78} \cite{Zw82} \cite{Zw89} \cite{DZ89}\ \cite{Zwa96}
\cite{Va92}; two nice reviews are \cite{SS05} \cite{EPZ04}). This possibility
is consistent with the usual perturbative point of view since, in the case of
$SU(N)$ Yang-Mills theories, for a "small enough" potential $A_{\mu}$ (with
respect to a suitable functional norm \cite{Va92}), there are no zero-mode of
the FP operator in the Landau or Coulomb gauge.

It is also worth to emphasize\ that the issue of gauge fixing ambiguities
cannot be ignored in any case. In particular, even if gauge fixing choices
free of Gribov ambiguities can be found, still the presence of Gribov
ambiguities in other gauges gives rise to a breaking of the BRS symmetry at a
non-perturbative level (see, for instance, \cite{Fuj} \cite{Sor1} \cite{Sor2}
\cite{BaSo}).

Abelian gauge theories on flat space-time, are devoid of this problem, since
the Gribov copy equation for the smooth gauge parameter $\phi$ is
\begin{equation}
\partial_{i}\partial^{i}\phi=0\;\;\;\mathrm{or}\;\;\;\partial_{\mu}%
\partial^{\mu}\phi=0 \label{abelian}%
\end{equation}
which on flat space-time (once the time coordinate has been Wick-rotated:
$t\rightarrow i\tau$) has no smooth non-trivial solutions fulfilling the
physical boundary conditions. In fact, the situation changes dramatically when
we consider an Abelian gauge field propagating on a curved background: it was
shown in \cite{CGO} that, quite generically, a proper gauge fixing in the
Abelian case cannot be achieved. Furthermore, it has been recently pointed out
\cite{ACGO} that, at least in the case of gravitational theories in\ 2+1
dimensions, gauge fixing ambiguities may provide one with a valuable tool to
achieve SUSY breaking.

For these reasons, the issue of the Gribov copies in the case of non-Abelian
gauge theories on curved spaces as well as on spaces with non-trivial
topologies is of interest. In many physically relevant situations (such as
close to a black hole, in neutron stars and even more in quarks and hybrid
star \cite{QUARKSTAR} and in cosmological setups) the curved nature of
space-time cannot be ignored. Thus, in those situations it is important to
consider the dynamics of QCD on a curved background. In the present paper we
will analyze the issue of the appearance of Gribov copies by analyzing the
curved generalization of the Gribov pendulum equation in the Coulomb
gauge\footnote{We will consider the Coulomb instead of the Landau gauge in
order to avoid the subtelties related to the Wick rotation on curved
spacetimes.}. Here we will consider the class of static curved spacetimes with
spherical symmetry as backgrounds. We will construct necessary conditions in
order for spherically symmetric backgrounds to admit solutions of the Gribov
pendulum equation representing copies of the vacuum and satisfying the strong
boundary conditions. We will show with explicit examples that the curvature of
the spacetime can generate quite non-trivial deformations of the Gribov horizon.

\bigskip

The paper is organized as follows. In section two, the curved generalization
of the Gribov pendulum in the Coulomb gauge will be constructed, and the
strong and weak boundary conditions will be given. In section three we analyze
the existence of copies in the background corresponding to AdS spacetime. In
the fourth section it will be shown that smooth solutions of the Gribov
pendulum equation exist and the analysis of the boundary conditions in terms
of an effective Schr\"{o}dinger equation will be also discussed. In the fifth
section, background metrics admitting copies of the vacuum satisfying the
strong boundary condition will be constructed. Some conclusions will be drawn
in the last section

\section{Curved Gribov pendulum}

The main goal of the present paper is to analyze the new features of Gribov
ambiguities in the Coulomb gauge on a curved spherically symmetric background.
The metric of the curved backgrounds which will be considered here is%
\begin{equation}
ds^{2}=-g^{2}(r)dt^{2}+f^{2}(r)dr^{2}+r^{2}\left(  d\theta^{2}+\sin\theta
^{2}d\phi^{2}\right)  \ . \label{Metrspher}%
\end{equation}
The Coulomb gauge condition on the non-Abelian gauge potential $A_{\mu}^{a}$
of the $SU(2)$ gauge group reads
\begin{equation}
A_{0}^{a}=0\ ;\ \nabla^{i}A_{i}^{a}=0\ , \label{coulombgf}%
\end{equation}
where the spatial indices correspond to $i=1,2,3$ and the $\nabla^{i}$ stands
for the Levi-Civita connection of the metric (\ref{Metrspher}), with spatial
indices. It is easy to see that due to the form of the metric (\ref{Metrspher}%
), the Coulomb gauge condition transforms covariantly with respect to the
three dimensional spatial metric $ds_{\Sigma}^{2}$ of $t=const$ surfaces%
\begin{equation}
ds_{\Sigma}^{2}=\left(  g_{\Sigma}\right)  _{ij}dx^{i}dx^{j}:=f^{2}%
(r)dr^{2}+r^{2}d\Omega^{2}\ , \label{indumetr}%
\end{equation}
where $d\Omega$ stands for the line element of the two sphere. The gauge
fixing (\ref{coulombgf}) then can be written as:%
\begin{equation}
\ \nabla^{i}A_{i}^{a}=\frac{1}{\sqrt{\det g_{\Sigma}}}\partial_{j}\left(
\sqrt{\det g_{\Sigma}}g_{\Sigma}^{ji}A_{i}^{a}\right)  =0\ . \label{cougauge}%
\end{equation}
Let us consider an element of the gauge group of the following form:
\begin{equation}
U\left(  x^{\mu}\right)  =\exp\left(  i\frac{\alpha\left(  r\right)  }{2}%
x^{i}\sigma_{i}\right)  \label{gaugtransf}%
\end{equation}
where $\sigma_{i}$ are the flat Pauli matrices and $x^{i}$ is a normalized
radial contravariant vector on $g_{\Sigma}$, which in the above coordinate
system reads%
\[
\overrightarrow{x}=\left(  \sin\theta\cos\phi,\sin\theta\sin\phi,\cos
\theta\right)  \ ,
\]
so that%
\begin{equation}
\mathbf{I=}\left(  x^{i}\sigma_{i}\right)  \left(  x^{j}\sigma_{j}\right)  \ ,
\end{equation}
$\mathbf{I}$ being the $2\times2$ identity matrix. It can be seen that
$U^{\dag}=U^{-1}$. Let us consider a background gauge potential $A_{i}^{\ast}$
of the following form:%
\begin{equation}
A_{i}^{\ast}=i\varepsilon_{ijk}\frac{x^{j}\sigma^{k}}{r^{2}}\varphi(r)\ ,
\label{backgauge}%
\end{equation}
where $\varepsilon_{ijk}$ is the three-dimensional Levi-Civita tensor. Note
that $A_{i}^{\ast}$\ is divergence free for any radial function $\varphi(r)$:%
\begin{equation}
\nabla^{i}A_{i}^{\ast}=0\ ,\ \ \forall\ \varphi(r)\ .
\end{equation}
We will choose the above gauge potential in Eq. (\ref{backgauge}). Even if
this is not the most general transverse potential, we choose it because it
discloses very clearly the differences between the solutions of the Gribov
pendulum equation on flat and curved spacetimes\footnote{The more general
ansatz (see, for instance, \cite{SS05}) for the background gauge potential
which gives rise to a spherically symmetric Gribov pendulum equation does not
add new qualitative features}. We are now in position to derive the curved
generalization of the Gribov pendulum equation. One has to ask then for a
gauge transformation of the non-Abelian gauge potential $A_{i}^{\ast}$ in Eq.
(\ref{backgauge}) generated by the group element $U$ in Eq. (\ref{gaugtransf})
satisfying%
\begin{equation}
\nabla^{i}\left(  U^{-1}A_{i}^{\ast}U+U^{-1}\partial_{i}U\right)
=0\ \label{prependulum}%
\end{equation}
The existence of solutions for the equation above is a necessary condition for
the appearance of Gribov copies of $A_{i}^{\ast}$. Explicitly on the
spherically symmetric spacetimes we are considering the gauge fixing equations
imply that%
\begin{equation}
\left(  \frac{r^{2}\alpha^{\prime}}{f}\right)  ^{\prime}=2f\left(
1-2\varphi\right)  \sin\alpha\ , \label{pendulum}%
\end{equation}
where primes denote derivation with respect to the radial coordinate. Note
that this equations is invariant under the transformation $f\rightarrow Cf$
and $\left(  1-2\varphi\right)  \rightarrow C^{-2}\left(  1-2\varphi\right)
$, provided $C$ is a constant.

\subsection{Strong and weak boundary conditions on curved spaces}

Here, we will discuss the weak and strong boundary conditions for the function
$\alpha$ in Schwarzschild-like coordinates as in Eq. (\ref{Metrspher}). The
importance to distinguish carefully copies satisfying strong and weak boundary
conditions comes from the following fact (well known in the flat case). When
the solution of the Gribov pendulum equation satisfies the weak boundary
conditions%
\begin{equation}
\alpha\underset{r\rightarrow\infty}{\rightarrow}(2n+1)\pi+O(1/r^{\eta
})\ ,\ \eta>0\ ,
\end{equation}
the corresponding copy%
\begin{equation}
U\left(  x^{\mu}\right)  =\exp\left(  i\frac{\alpha\left(  r\right)  }{2}%
x^{j}\sigma_{j}\right)  =\mathbf{1}\cos\left(  \frac{\alpha\left(  r\right)
}{2}\right)  +ix^{j}\sigma_{j}\sin\left(  \frac{\alpha\left(  r\right)  }%
{2}\right)  \ ,
\end{equation}
does not approach to an element of the center of the gauge group at spatial
infinity. A copy of this type it is not problematic since it can be discarded
with the argument that it changes the definition of (non-Abelian) charge at
infinity and so it does not give rise to a proper gauge transformation (see
\cite{Benguria:1976in} \cite{Gri78}).

On the other hand, when a solution of the Gribov pendulum equation satisfies
the strong boundary conditions, the corresponding copy does approach to an
element of the center of the gauge group at spatial infinity. A copy of this
type is particularly problematic since it belongs to the class of proper gauge
transformations and would represent a failure of the whole gauge fixing
procedure. Of course, this was one of the main arguments behind the
Gribov-Zwanziger idea of "cutting" the path integral when the first copies
satisfying the strong boundary conditions appear. Indeed, the worst case would
be to have a copy of the vacuum $A_{\mu}=0$ fulfilling the strong boundary
conditions since, in this case, not even usual perturbation theory leading to
the standard Feynman rules in the Landau or Coulomb gauge would be well
defined. In QCD on flat space this does not happen but we will show here that
whenever the theory is considered in a curved background, the situation
becomes much more delicate.

\bigskip

- \textbf{Weak boundary conditions: }The weak boundary condition for a copy on
the metric (\ref{Metrspher}), corresponds to look for a solution of the curved
Gribov pendulum equation Eq. (\ref{pendulum}) which behaves as%
\begin{align}
\alpha\underset{r\rightarrow\infty}{\rightarrow}(2n+1)\pi+O(1/r^{\eta
})\ ,\ \eta &  >0\ ,\ \ \varphi\underset{r\rightarrow\infty}{\rightarrow
}const+o(1/r)\ ,\nonumber\\
\alpha\underset{r\rightarrow0}{\rightarrow}2m\pi+O(r^{\gamma})\ ,\ \gamma &
>0\ ,\ \ \ m,n\in%
\mathbb{Z}
\ . \label{origin}%
\end{align}
As it occurs on flat spacetime \cite{SS05}, as far as the behavior of the
solution $\alpha$ close to the origin is concerned, both in the case of weak
and in the case of strong boundary conditions one has to require that the
condition in Eq. (\ref{origin}) holds otherwise the copy generated by the
solution $\alpha$ would not be regular at the origin. As it will be discussed
in the next sections, in the case in which a star is considered as a
gravitational background, the situation is quite different.

Since the Christoffell symbols do not enter directly in the expression
$U^{-1}A_{\mu}U+U^{-1}\partial_{\mu}U$ (since $U$ behaves as a scalar under
diffeomorphisms), in terms of $\alpha$ both the strong and the weak boundary
conditions keep forms similar to the corresponding flat cases. In particular,
this implies that also on a spherically symmetric curved space as in Eq.
(\ref{Metrspher}) the gauge transformation generated by an element of the
group of the form $U\left(  x^{\mu}\right)  =\exp\left(  i\frac{\alpha\left(
r\right)  }{2}x^{i}\sigma_{i}\right)  $ will change the definition of
non-Abelian charge as a surface integral at spatial infinity unless
$\alpha(r)$ approaches to $2n\pi$ (strong boundary condition).

On flat spacetime, a vector potential $A_{\mu}^{C}$ which generates a
Coulomb-like electric field decays as%
\begin{equation}
A_{\mu}^{C}\underset{r\rightarrow\infty}{\approx}\frac{1}{r}+O\left(  \frac
{1}{r^{p}}\right)  \ \text{with }p>1\ ,
\end{equation}
in order for the corresponding electric field to decay as $1/r^{2}$. On AdS
spacetime, the metric function $f$ is given in Eq. (\ref{AdSgrr}). Therefore,
the electric field has to decay also as $1/r^{2}$ in order to generate a
finite charge. One can see this as follows: the electric (as well as the
non-Abelian) charge can be written in this way
\begin{equation}
Q^{\left(  a\right)  }=-\int_{\partial\Sigma}d^{2}x\sqrt{\gamma^{\partial
\Sigma}}n_{\mu}s_{\nu}F^{\left(  a\right)  \mu\nu}\ ,
\end{equation}
where $a$ is in the adjoint representation of $su\left(  2\right)  $,
$\partial\Sigma$ is the boundary of the spacelike section $\Sigma$, with
induced metric $\gamma^{\partial\Sigma}$, $n_{\mu}$ is a normalized future
pointing timelike vector ($n_{\mu}n^{\mu}=-1$) and $s_{\mu}$ is normal to
$\partial\Sigma$ and normalized as $s_{\mu}s^{\mu}=1$. Thus, in the AdS case,
in order to have a finite charge the electric field has to decay as $1/r^{2}$
and correspondingly the vector potential generating an electric field has to
decay as $1/r$.

Therefore, as it happens on flat space-times (see, for instance,
\cite{Gri78}), in the case of the weak boundary conditions, one has to require
the function $\varphi(r)$ appearing in the ansatz for the transverse vector
potential in Eq. (\ref{backgauge}), to decay as%
\begin{equation}
\varphi(r)\underset{r\rightarrow\infty}{\approx}const+O(1/r)\ .
\label{adsfalloffweak}%
\end{equation}

- \textbf{Strong boundary conditions: }The strong boundary condition on the
metric (\ref{Metrspher}) given in Schwarzschild-like coordinates corresponds
to ask that when $r\rightarrow\infty$, the solution of the curved Gribov
pendulum equation Eq. (\ref{pendulum}) behaves as%
\begin{equation}
\alpha\underset{r\rightarrow\infty}{\rightarrow}2n\pi+O(1/r^{\eta
})\ ,\ \varphi(r)\underset{r\rightarrow\infty}{\approx}\frac{1}{r^{\varepsilon
}}\ ,\ \ \eta\ ,\ \varepsilon>0\ , \label{adsradialstrong}%
\end{equation}
while close to the origin the condition in Eq. (\ref{origin}) must hold in
order the copy generated by $\alpha$ to be regular.

At a first glance, the curved Gribov pendulum in vacuum (which corresponds to
Eq. (\ref{pendulum}) with $\varphi=0$) could look like a flat Gribov pendulum
(in which case $f=1$) in a non-trivial background gauge field. If this would
be the case, then it would also be easy to construct examples of curved
background supporting copies of the vacuum satisfying the strong boundary
conditions. Obviously, a non-Abelian gauge theory on a curved background
supporting copies of the vacuum satisfying the strong boundary conditions
would be pathological. However in many important cases (such as constant
curvature backgrounds and spherically symmetric black hole spacetimes), such a
resemblance is misleading. As it will be shown in the next sections, in these
cases solutions of the curved Gribov pendulum equation representing copies of
the vacuum satisfying the strong boundary conditions cannot be constructed.

\bigskip

Defining $\tau=\tau\left(  r\right)  $ by%
\begin{equation}
\tau^{\prime}=\frac{\partial\tau}{\partial r}=\frac{f}{r^{2}}\ , \label{jacob}%
\end{equation}
the curved Gribov pendulum equation (\ref{pendulum}) can be transformed in the
following useful form%
\begin{equation}
\frac{\partial^{2}\alpha}{\partial\tau^{2}}=2r^{2}\left(  1-2\varphi\right)
\sin\alpha\ , \label{taupendulum}%
\end{equation}
where the variable $r$ has to be expressed in terms of $\tau$ using Eq.
(\ref{jacob}). As it will be discussed in the next section, this form of the
equation allows to deal in a very effective way with the problem of imposing
strong and weak boundary conditions.

We will first focus on the case of AdS as a background metric. According to
the AdS/CFT correspondence, it is possible to explore the non-perturbative
regime of supersymmetric Yang-Mills theories by performing semiclassical
computations in the bulk of asymptotically AdS background \cite{maldacena}
\cite{magoo}. Recently, this correspondence has also been extended to the
context of condensed matter physics (see for two recent reviews
\cite{adscondmat}).

\bigskip

The metric (\ref{Metrspher}) reduces to the metric on AdS spacetime, provided%
\begin{equation}
f\left(  r\right)  =\frac{1}{\sqrt{1+\frac{r^{2}}{l^{2}}}}\ ; \label{AdSgrr}%
\end{equation}
where $l$ is the AdS curvature. Eq. (\ref{jacob}) implies that%
\begin{equation}
\tau=-\frac{\sqrt{1+\frac{r^{2}}{l^{2}}}}{r}\ \Rightarrow r^{2}=\frac
{1}{\left(  l\tau\right)  ^{2}-1}\ ,\ \tau<0\ . \label{adsgrr2}%
\end{equation}
In particular at spatial infinity we have%
\[
r\rightarrow+\infty\ \Leftrightarrow\ l\tau\rightarrow-1^{-}\ .
\]
Therefore, in the AdS case the Gribov pendulum equation (\ref{taupendulum})
can be rewritten as%
\begin{align}
\frac{\partial^{2}\alpha}{\partial\tau^{2}}  &  =V_{AdS}\left(  \tau\right)
\sin\alpha\ .,\label{adspendulum}\\
V_{AdS}\left(  \tau\right)   &  :=\frac{2\left(  1-2\varphi\right)  }{\left(
l\tau\right)  ^{2}-1}\ . \label{effectiveelasticonstant}%
\end{align}
On the other hand, on flat spacetime $f\left(  r\right)  =1$ so that
$\tau=-\frac{1}{r}$ and%
\begin{equation}
r\rightarrow+\infty\ \Leftrightarrow\ \tau\rightarrow0^{-}\ .
\end{equation}
Eq. (\ref{pendulum}), then reduces to the flat Gribov pendulum equation which
reads%
\begin{align}
\frac{\partial^{2}\alpha}{\partial\tau^{2}}  &  =V_{flat}\left(  \tau\right)
\sin\alpha\ ,\label{flatpendulum}\\
V_{flat}\left(  \tau\right)   &  :=\frac{2\left(  1-2\varphi\right)  }%
{\tau^{2}}\ . \label{effectiveelastconstantflat}%
\end{align}
It is worth emphasizing here that, when one writes the curved Gribov pendulum
equation in terms of the coordinate $\tau$, the main difference between the
AdS and the flat cases occurs close to the singularities of the effective
potentials\footnote{The reason to call $V_{AdS}(\tau)$ and $V_{flat}(\tau)$
effective potentials will be manifest in the next sections.} $V_{AdS}(\tau)$
and $V_{flat}(\tau)$ appearing in Eqs. (\ref{effectiveelasticonstant}) and
(\ref{effectiveelastconstantflat}). Close to the singularity (when
$l\tau\rightarrow-1^{-}$), the effective AdS potential $V_{AdS}(\tau)$
diverges as $1/\tau$ while the effective potential $V_{flat}(\tau)$
corresponding to the flat metric diverges when $\tau\rightarrow0$ as
$1/\tau^{2}$. On the other hand, as soon as one moves away from the
corresponding singularities (namely, when $\left\vert \tau\right\vert >1$),
the Gribov pendulum equations in the AdS and flat cases look the same.

The curved Gribov pendulum equation corresponding to the Coulomb gauge on the
spherically symmetric background in Eq. (\ref{Metrspher}) can be derived as
the Euler-Lagrange equation of the following functional:%
\begin{equation}
N\left[  \alpha\right]  =\int\sqrt{\det g_{\Sigma}}d^{3}xTr\left[  \left(
U^{-1}A_{i}^{\ast}U+U^{-1}\partial_{i}U\right)  ^{2}\right]  \ . \label{norm0}%
\end{equation}
When one inserts into the above expression Eqs. (\ref{gaugtransf}) and
(\ref{backgauge}) one gets the following useful expression (which reduces to
the known flat case \cite{Gri78} when $f=1$):
\begin{equation}
N\left[  \alpha\right]  =\int_{a}^{\infty}\frac{dr}{f}\left\{  \left(
r\alpha^{\prime}\right)  ^{2}+8f^{2}\left(  1-2\varphi\right)  \left[
1-\cos^{2}\left(  \frac{\alpha}{2}\right)  \right]  \right\}  \ .
\label{Normgri1}%
\end{equation}
\bigskip

The main goal of this paper is to show that the solutions of the curved Gribov
pendulum equation on curved spacetime can behave in a totally different manner
with respect to the flat case. In particular, we will show that there are many
physically interesting curved backgrounds that may admit copies of the vacuum
satisfying the strong boundary condition (to be defined in the next section).
If one accepts the interpretation of \cite{Gri78} \cite{Zw82} \cite{Zw89}
\cite{DZ89}\ \cite{Zwa96} \cite{Sor1} \cite{Sor2}, our results would imply
that the infrared structure of QCD on curved spacetimes could be quite
different from the infrared structure on flat space-times.

\section{On the existence and uniqueness of solutions on AdS}

In this section we will describe the mathematical technique, based on the
contraction theorem, which allows to prove existence and uniqueness of
non-linear Gribov pendulum equations for $r$ larger than a suitable critical
radius (defined below). When $r$ is small, provided%
\begin{equation}
f^{2}\left(  r\right)  =1+O\left(  r^{2}\right)  \ ,
\end{equation}
the metric approaches to flat metric and, for globally flat background
metrics, the issue of existence and uniqueness of solutions of the Gribov
pendulum equation is well understood. Moreover, the most interesting technical
differences in the procedure with respect to the flat case when AdS or
asymptotically AdS spacetime are considered as backgrounds, appear for $r$
larger than a critical radius (see also the comments after Eq.
(\ref{effectiveelastconstantflat})). Thus, we will focus on the analysis of
the problem for $r$ larger than a critical radius defined below. The goal of
this section is to provide one with a rigorous justification of the effective
Schr\"{o}dinger approach to the analysis of the weak and strong boundary
conditions, which is useful on curved backgrounds approaching AdS in the
asymptotic region.

The statement of the theorem (\cite{berger} \cite{giltrud}) is the following:

Let $S$ a complete metric (Banach) space. A metric space is a space in which a
distance $d\left(  X,Y\right)  $ between any pair of elements of the space is
defined
\begin{equation}
d\left(  X,Y\right)  \in%
\mathbb{R}
\ ,\ \ X,Y\in S\ .
\end{equation}
Complete metric space means that, with respect to the metric, from every
Cauchy sequence one can extract a convergent subsequence (see, for instance,
\cite{berger}). Let $T$ be a map from the metric space $S$ into itself:%
\begin{equation}
T\left[  .\right]  :S\rightarrow S\ .
\end{equation}
If the map $T\left[  .\right]  $ is a contraction, namely for all $X\in S$ and
$Y\in S$%
\begin{equation}
d\left(  T\left[  X\right]  ,T\left[  Y\right]  \right)  \leq Md\left(
X,Y\right)  \ ,\ \text{with}\ \ M<1,
\end{equation}
then the map $T\left[  .\right]  $ has only one fixed point. In other words,
if the map $T\left[  .\right]  $ is a contraction of a complete metric space
then there exist a unique solution to the equation%
\begin{equation}
T\left[  X\right]  =X\ .
\end{equation}
Hereafter we will focus on the asymptotic region, defined by $r\rightarrow
\infty$ (in a precise sense as it will be explained in a moment).

The idea is to write the non-linear equation one is interested in (Eq.
(\ref{pendulum}) in our case) in the form of a fixed point equation for a
suitable map and then, try to prove that the map is a contraction for some
complete metric space. Let us define the following operator $T_{\varphi}$:%
\begin{equation}
T_{\varphi}\left[  \alpha\right]  \left(  r\right)  \equiv A+B\int_{r^{\ast}%
}^{r}\frac{ds}{s^{2}\sqrt{1+s^{2}}}+\int_{r^{\ast}}^{r}\frac{1}{s^{2}%
\sqrt{1+s^{2}}}\left[  \int_{r^{\ast}}^{s}\frac{2\left(  1-2\varphi\left(
\rho\right)  \right)  \sin\alpha\left(  \rho\right)  }{\sqrt{1+\rho^{2}}}%
d\rho\right]  ds\ . \label{fixedpoint}%
\end{equation}
where $A$ and $B$ are arbitrary constants, and $r^{\ast}$ defines a critical
radius. It is easy to see that the Gribov pendulum equation in Eq.
(\ref{pendulum}) in the AdS case can be written as a fixed point equation for
the operator defined in Eq. (\ref{fixedpoint}), i.e., if $\widetilde{\alpha}$
is a fixed point of $T_{\varphi}$%
\begin{equation}
T_{\varphi}\left[  \widetilde{\alpha}\right]  \left(  r\right)  \equiv
\widetilde{\alpha}\left(  r\right)  \label{fixedpointform}%
\end{equation}
then the same $\widetilde{\alpha}$ is a solution of Eq. (\ref{pendulum}) with
AdS as a background geometry. This can be seen directly by applying
consecutively two derivatives at the right hand side of Eq. (\ref{fixedpoint}%
). Note that we have fixed the AdS radius $l$ to $1$. Thus, we will prove that
the above operator has a fixed point by using the above mentioned theorem. Let
us define $S$ as the space of functions which are continuos and bounded on
$\left[  r^{\ast},\ \infty\right[  $, i.e.
\begin{equation}
S\equiv\left\{  \left.  \alpha\right\vert \ \alpha\in C\left[  r^{\ast
},\ \infty\right[  \ ,\ \ \left\vert \alpha(r)\right\vert <M_{\alpha
}\ \ \forall\ r\in\left[  r^{\ast},\ \infty\right[  \right\}  \ .
\label{funspace}%
\end{equation}
The radius $r^{\ast}$ will be determined in a moment and the constants $A$ and
$B$ correspond to the value of $\widetilde{\alpha}$ and its derivative at
$r^{\ast}$, respectively:
\begin{equation}
\widetilde{\alpha}(r^{\ast})=A\ ,\ \ \widetilde{\alpha}^{\prime}(r^{\ast
})=\frac{B}{\left(  r^{\ast}\right)  ^{2}\sqrt{1+\left(  r^{\ast}\right)
^{2}}}\ .
\end{equation}
This functional space is a Banach space (see, for instance, \cite{berger})
with respect to the following distance $d\left(  \alpha,\beta\right)  $:%
\begin{equation}
d\left(  \alpha,\beta\right)  =\underset{r\in\left[  r^{\ast},\ \infty\right[
}{\sup}\left\vert \alpha(r)-\beta(r)\right\vert \ . \label{meba}%
\end{equation}
It is easy to see that the operator $T_{\varphi}$ defined on (\ref{fixedpoint}%
), maps $S_{A,B}$ into itself since%
\begin{equation}
\left\vert \alpha(r)\right\vert <M_{\alpha}\ \Rightarrow\ \left\vert
T_{\varphi}\left[  \alpha\right]  \left(  r\right)  \right\vert <\widetilde
{M}_{\alpha}\ ,
\end{equation}
where $\widetilde{M}_{\alpha}$ may be different from $M_{\alpha}$. Indeed, let
us consider the following function $I_{1}(r)$:%
\begin{align}
I_{1}(r)  &  =\int_{r^{\ast}}^{r}\frac{ds}{s^{2}\sqrt{1+s^{2}}}\ ,\\
\left\vert I_{1}(r)\right\vert  &  \leq\frac{1}{r^{\ast}}\ \ \forall\ r\ .
\end{align}
Hence one has%
\begin{align}
\left\vert T_{\varphi}\left[  \alpha\right]  \left(  r\right)  \right\vert  &
\leq\left\vert A\right\vert +\frac{\left\vert B\right\vert }{r^{\ast}}%
+\int_{r^{\ast}}^{r}\frac{ds}{s^{2}\sqrt{1+s^{2}}}\left[  \int_{r^{\ast}}%
^{s}\frac{2\left\vert \left(  1-2\varphi\left(  \rho\right)  \right)
\right\vert \left\vert \sin\alpha\left(  \rho\right)  \right\vert }%
{\sqrt{1+\rho^{2}}}d\rho\right]  \leq\\
&  \leq\left\vert A\right\vert +\frac{\left\vert B\right\vert }{r^{\ast}%
}+\left\vert 1+2M_{\varphi}\right\vert \int_{r^{\ast}}^{r}\frac{ds}{s^{2}%
\sqrt{1+s^{2}}}\left[  \int_{r^{\ast}}^{s}\frac{2}{\sqrt{1+\rho^{2}}}%
d\rho\right]  <\\
&  <\left\vert A\right\vert +\frac{\left\vert B\right\vert }{r^{\ast}%
}+\left\vert 1+2M_{\varphi}\right\vert \int_{r^{\ast}}^{r}\frac{2\left(
s-r^{\ast}\right)  ds}{s^{2}\sqrt{1+s^{2}}}<\\
&  <\left\vert A\right\vert +\frac{\left\vert B\right\vert }{r^{\ast}%
}+2\left\vert 1+2M_{\varphi}\right\vert \int_{r^{\ast}}^{r}\frac{ds}%
{s\sqrt{1+s^{2}}}<+\infty\ ,
\end{align}
where we used the fact that $\varphi$ in Eq. (\ref{backgauge}) is bounded and
smooth everywhere so that
\begin{align}
\left\vert \left(  1-2\varphi\left(  r\right)  \right)  \right\vert  &
<\left\vert 1+2M_{\varphi}\right\vert \ \ \forall\ r\ ,\\
M_{\varphi}  &  =\underset{r\in\left[  0,\ \infty\right[  }{\sup}\left\vert
\varphi\left(  r\right)  \right\vert <\infty\ .
\end{align}
We will show that it is possible to choose the radius $r^{\ast}$ such that the
operator $T_{\varphi}$ is a contraction of the Banach space in Eq.
(\ref{funspace}) with the distance in Eq. (\ref{meba}). To see this, one has
to compute $\left\vert T_{\varphi}\left[  \alpha\right]  -T_{\varphi}\left[
\beta\right]  \right\vert $ where $\alpha$, $\beta\in S$:%
\begin{align}
\left\vert T_{\varphi}\left[  \alpha\right]  -T_{\varphi}\left[  \beta\right]
\right\vert  &  \leq\int_{r^{\ast}}^{r}\frac{ds}{s^{2}\sqrt{1+s^{2}}}\left[
\int_{r^{\ast}}^{s}\frac{2\left\vert \left(  1-2\varphi\left(  \rho\right)
\right)  \right\vert \left\vert \sin\alpha\left(  \rho\right)  -\sin
\beta(r)\right\vert }{\sqrt{1+\rho^{2}}}d\rho\right]  \leq\\
&  \leq c\left(  \sup\left\vert \alpha(r)-\beta(r)\right\vert \right)
\left\vert 1+2M_{\varphi}\right\vert \int_{r^{\ast}}^{r}\frac{ds}{s^{2}%
\sqrt{1+s^{2}}}\left[  \int_{r^{\ast}}^{s}\frac{2}{\sqrt{1+\rho^{2}}}%
d\rho\right]  <\\
&  <c\left(  \left\vert 1+2M_{\varphi}\right\vert \right)  d\left(
\alpha,\beta\right)  \int_{r^{\ast}}^{r}\frac{2\left(  s-r^{\ast}\right)
ds}{s^{2}\sqrt{1+s^{2}}}<\\
&  <2c\left(  \left\vert 1+2M_{\varphi}\right\vert \right)  d\left(
\alpha,\beta\right)  \int_{r^{\ast}}^{r}\frac{ds}{s^{2}}<\frac{2c\left(
\left\vert 1+2M_{\varphi}\right\vert \right)  }{r^{\ast}}d\left(  \alpha
,\beta\right)  \ \Rightarrow\label{lasine}\\
\sup\left\vert T_{\varphi}\left[  \alpha\right]  -T_{\varphi}\left[
\beta\right]  \right\vert  &  =d\left(  T_{\varphi}\left[  \alpha\right]
,T_{\varphi}\left[  \beta\right]  \right)  <\frac{2c\left(  \left\vert
1+2M_{\varphi}\right\vert \right)  }{r^{\ast}}d\left(  \alpha,\beta\right)
\label{lasine2}%
\end{align}
where we have used the trigonometric identity%
\begin{equation}
\sin\alpha-\sin\beta=2\cos\left(  \frac{\alpha+\beta}{2}\right)  \sin\left(
\frac{\alpha-\beta}{2}\right)  \ ,
\end{equation}
as well as the inequalities%
\begin{equation}
\left\vert \sin x\right\vert \leq\left\vert x\right\vert \ ,.\left\vert \cos
x\right\vert \leq1\ ,\ \ \ \forall\ x\ .
\end{equation}
Eqs. (\ref{lasine}) and (\ref{lasine2}) show that a sufficient condition in
order for $T_{\varphi}$ to be a contraction is to choose $r^{\ast}$ such that:%
\begin{equation}
\frac{2\left\vert 1+2M_{\varphi}\right\vert }{r^{\ast}}<1\ . \label{lastheor}%
\end{equation}
Thus, if one chooses $r^{\ast}$ satisfying the inequality in Eq.
(\ref{lastheor}) then Eq. (\ref{pendulum}) has a unique solution in the AdS
case. It is worth to note that in all the previous steps the presence of the
curved metric (through the AdS factor $1/\sqrt{1+r^{2}}$) helped in obtaining
the required bounds. Of course, if the constant $A$ is chosen to be a multiple
of $\pi$ and $B$ vanishes then, because of the above result, the unique
solution is the constant, i.e.%
\begin{equation}
A=n\pi\ \vee\ B=0\ \Rightarrow\ \alpha(r)=n\pi\ \ \forall\ r\geq r^{\ast}\ .
\end{equation}
One can observe that the solution is at least $C^{2}\left[  r^{\ast}%
,\infty\right[  $ since, as the integral form of the equation shows, one can
take at least two derivatives. Furthermore, not only the solution but also the
first and the second derivative of the solution are bounded as one can deduce
from the equation in the "fixed point" form in Eqs. (\ref{fixedpoint}) and
(\ref{fixedpointform}). This implies that, necessarily, one has%
\[
\alpha(r)\underset{r\rightarrow\infty}{\rightarrow}n\pi\ ,
\]
otherwise the second derivative would not be bounded.

\subsection{Schr\"{o}dinger equation approach}

In order to analyze the issue of existence of copies satisfying strong
boundary conditions, one can use an effective one-dimensional Schr\"{o}dinger
equation. In the AdS case, it is useful to consider the equation with the
change of coordinate in Eqs. (\ref{jacob}) and (\ref{adsgrr2}), and the
corresponding Gribov pendulum equation in Eqs. (\ref{adspendulum}) and
(\ref{effectiveelasticonstant}). Because of the theorem discussed in the
previous section, we know that bounded smooth solutions exist when
$r\rightarrow\infty\ $($l\tau\rightarrow-1^{-}$). Therefore, when%
\begin{equation}
V\left(  \tau\right)  \underset{l\tau\rightarrow-1^{-}}{\rightarrow}\infty\ ,
\end{equation}
in order for the solution to be bounded (taking into account that both the
first and the second derivatives of the solution must be bounded as well as):%
\begin{align}
\alpha\underset{r\rightarrow\infty}{\rightarrow}n\pi+O(1/r)\ \  &
\Leftrightarrow\label{asymptoti1}\\
&  \alpha\underset{l\tau+1\rightarrow0^{-}}{\rightarrow}n\pi+O\left(
l\tau+1\right)  \ . \label{asymptotic}%
\end{align}
Consequently the following, leading order approximation is justified
\begin{equation}
\sin\alpha\underset{l\tau+1\rightarrow0^{-}}{\approx}\ \left(  -1\right)
^{n}\alpha\ ,
\end{equation}
where $n$ odd (even) corresponds to the weak (strong) boundary conditions. For
these reasons, one is allowed to approximate for $r\gg r^{\ast}$ Eq.
(\ref{adspendulum}) as follows%
\begin{equation}
\frac{\partial^{2}\alpha}{\partial\tau^{2}}=\left(  -1\right)  ^{n}V\left(
\tau\right)  \alpha=W(\tau)\alpha\label{sesseas}%
\end{equation}
which can be analyzed as a Schr\"{o}dinger-like equation:%
\begin{align}
-u^{\prime\prime}+W(\tau)u  &  =Eu\ ,\ \ \ \tau\in\left]  -\infty,-\frac{1}%
{l}\right[ \label{schrodingerlike}\\
W(\tau)  &  =\left(  -\right)  ^{n}V\left(  \tau\right)  \ ,\ \ E=0.
\label{effectivepotential}%
\end{align}
Thus, the question of existence of normalizable copies reduces to the question
of existence of non-trivial normalizable eigenvectors (bound states) of the
above Schr\"{o}dinger-like problem with zero eigenvalue such that
\begin{equation}
u\underset{l\tau+1\rightarrow0^{-}}{\rightarrow}0\ .
\end{equation}
As far as the vacuum copies in AdS are concerned, $V\left(  \tau\right)  $ in
Eqs. (\ref{sesseas}) and (\ref{effectiveelasticonstant}) is always a positive
and monotone function in $\left]  -\infty,-1/l\right[  $ and diverges to
$+\infty$ when $l\tau+1\rightarrow0^{-}$ so that, in order for the effective
potential $W$ in Eq. (\ref{effectivepotential}) to have bound states, the only
possibility is that $n$ in Eq. (\ref{asymptoti1}) is an odd number. Namely, on
AdS, there are no vacuum copies satisfying the strong boundary
conditions\footnote{Note that it is always possible to choose the integration
constant such that the bound state vanishes in $l\tau+1$ as required by the
physics of the problem.}.

On the other hand, in order to have a copy satisfying the strong boundary
condition, it is enough as it happens on flat spacetimes, to consider
$\varphi$ in Eq. (\ref{sesseas}) which makes $V\left(  \tau\right)  $ negative
enough in order to produce a "valley" in the effective potential which
supports a bound state, even when $n$ in Eq. (\ref{asymptoti1}) is an even number.

It is worth to emphasize that this effective Schr\"{o}dinger approach when
applied to the flat case, in which the effective potential is (see Eqs.
(\ref{flatpendulum}) and (\ref{effectiveelastconstantflat}))
\begin{equation}
W(\tau)=\left(  -1\right)  ^{n}\frac{2\left(  1-2\varphi\right)  }{\tau^{2}%
}\ ,
\end{equation}
reproduces the well known results such as the absence of vacuum copies
satisfying the strong boundary conditions and also the need to have a factor
$1-2\varphi$ "negative enough" as to produce a valley supporting a non-trivial
bound state in a very intuitive manner. Furthermore, in this framework it is
quite apparent the difference between the asymptotic behaviors in the flat and
the AdS cases. The absolute value of the effective potential in the asymptotic
region in the AdS case\footnote{As mentioned before, in the AdS case the
asymptotic region $r\rightarrow\infty$ in the $\tau$ coordinate is defined as
$l\tau+1\rightarrow0^{-}$ while in the flat case the asymptotic region
$r\rightarrow\infty$ in the $\tau$ coordinate is defined as $\tau
\rightarrow0^{-}$.} diverges as $1/\tau$ while in the flat case it diverges as
$1/\tau^{2}$.

\section{Metrics with copies of the vacuum satisfying the strong boundary
condition}

In this section, we will describe sufficient conditions in order for the
spherically symmetric background metric in Eq. (\ref{Metrspher}) to support
copies of the vacuum. A background of this type would be quite pathological
and one may wonder whether if, at least in a semiclassical approach to quantum
gravity, backgrounds admitting copies of the vacuum should be discarded. This
is a reasonable consistency criterion since one would like QCD perturbation
theory to be well defined. Indeed, according to the point of view in
\cite{Gri78} \cite{Zw82} \cite{Zw89} \cite{DZ89}\ \cite{Zwa96} \cite{Sor1}
\cite{Sor2}, on such background spacetimes allowing copies of the vacuum
satisfying the strong boundary conditions not even perturbation theory around
a vanishing gauge field would be well defined.

A simple method inspired by the well known work of Henyey \cite{heniey} to
deduce necessary conditions for the appearance of copies of the vacuum, is to
interpret Eq. (\ref{pendulum}) in the case in which the background gauge field
in Eq. (\ref{backgauge}) vanishes, as an equation for the metric function $f$
appearing in (\ref{Metrspher}), assuming that $\alpha$ is everywhere regular
and satisfies the strong boundary conditions when $r\rightarrow\infty$. In the
case in which the space-time is everywhere regular (the case of black hole and
stars will be considered in the next sub-sections) one can express $f$ in
terms of the copy $\alpha$ as follows:%
\begin{equation}
f(r)^{2}=\frac{\left(  r^{2}\partial_{r}\alpha\right)  ^{2}}{C+4\int_{0}%
^{r}x^{2}\left(  \partial_{x}\alpha\right)  \left(  \sin\alpha(x)\right)  dx}
\label{inverscatt1}%
\end{equation}
where $C$ is an integration constant\footnote{Replacing $f^{2}\rightarrow
f^{-1}$ this expression reduced to Eq. (22) of \cite{CGO} for $\sin\left(
\alpha\right)  \approx\alpha.$}. As mention before, in order for the metric to
be regular close to the origin one has to require that%
\begin{equation}
f(r)^{2}\underset{r\rightarrow0}{\approx}1+kr^{2}+O(r^{4})\ ,
\end{equation}
where $k$ is a real constant. By a direct expansion, one can see that this
implies that $C=0$ in Eq. (\ref{inverscatt1}). To fix the ideas, one can take
a function $\alpha$ increasing monotonically from $0$ to $2\pi$ at infinity:
\begin{align}
\alpha(0)  &  =0\ ,\ \alpha(r)\underset{r\rightarrow\infty}{\rightarrow}%
2\pi\ ,\label{ch1}\\
\ \partial_{r}\alpha &  >0\ \ \forall\ r>0\ ,\ \ \ \partial_{r}\alpha
\underset{r\rightarrow\infty}{\rightarrow}0\ . \label{ch2}%
\end{align}
With this choice the integrand in the denominator in Eq. (\ref{inverscatt1})
does not change sign for small $r$ but one would have $f(r)^{2}<0$ for $r$
large enough. To see this it is convenient to consider the following change of
variable in the integral in the denominator in Eq. (\ref{inverscatt1})%
\begin{align}
\partial_{x}\alpha\ dx  &  =d\alpha\ \Rightarrow\label{varint1}\\
\int_{0}^{r}x^{2}\left(  \partial_{x}\alpha\right)  \left(  \sin
\alpha(x)\right)  dx  &  =\int_{\alpha(0)}^{\alpha(r)}\left(  x\left(
\alpha\right)  \right)  ^{2}\sin\alpha d\alpha\ , \label{varint2}%
\end{align}
where $x(\alpha)$ is the inverse function of $\alpha(x)$ (which exist because
of our hypothesis). The integral in the denominator in Eq. (\ref{inverscatt1})
up to infinity reads%
\begin{align}
\int_{0}^{\infty}x^{2}\left(  \partial_{x}\alpha\right)  \left(  \sin
\alpha(x)\right)  dx  &  =\int_{0}^{2\pi}\left(  x\left(  \alpha\right)
\right)  ^{2}\sin\alpha d\alpha\label{integr1}\\
&  =\int_{0}^{\pi}\left(  x\left(  \alpha\right)  \right)  ^{2}\sin\alpha
d\alpha+\int_{\pi}^{2\pi}\left(  x\left(  \alpha\right)  \right)  ^{2}%
\sin\alpha d\alpha\ , \label{integr2}%
\end{align}
where%
\[
\int_{0}^{\pi}\left(  x\left(  \alpha\right)  \right)  ^{2}\sin\alpha
d\alpha>0\ ,\ \ \int_{\pi}^{2\pi}\left(  x\left(  \alpha\right)  \right)
^{2}\sin\alpha d\alpha<0\ .
\]
Since $\partial_{r}\alpha>0$ then $x(\alpha)$ is an increasing function of
$\alpha$, therefore the absolute value of the second integral on the right
hand side of Eq. (\ref{integr2}) is larger than the first\footnote{Note that
the two integrals in Eq. (\ref{integr2}) would be equal and opposite without
the factor $\left(  x(\alpha)\right)  ^{2}$. \ } and consequently the integral
in Eq. (\ref{integr1}) is negative. Thus, with the choice in Eqs. (\ref{ch1})
and (\ref{ch2}) for $r$ large enough $f^{2}$ is negative and solutions
$\alpha$ satisfying the strong boundary conditions cannot appear. By repeating
basically the same argument, one can see that the same conclusion would hold
for any choice in which $\alpha$ is monotone $\forall\ r>0$. It is worth to
point out that the situation does not change qualitatively if one chooses
$\alpha$ as a monotone decreasing function with $\alpha(0)=2\pi$ and
$\alpha(r)\underset{r\rightarrow\infty}{\rightarrow}0$.\newline

If one chooses a function $\alpha$ which is not monotone, then $\partial
_{r}\alpha$ vanishes at least once for $r^{\ast}>0$. Let us assume first that
$\partial_{r}\alpha$ vanishes just once at $r^{\ast}$, and that this is a
simple zero%
\begin{align}
\alpha(0)  &  =0\ ,\ \alpha(r)\underset{r\rightarrow\infty}{\rightarrow
}0\ ,\ \left.  \partial_{r}\alpha\right\vert _{r=r^{\ast}}=0\ ,\ \ \pi
<\alpha(r^{\ast})<2\pi\ ,\label{ch1.2}\\
\ \partial_{r}\alpha &  >0\ \ \forall\ r<r^{\ast}\ ,\ \partial_{r}%
\alpha<0\ \ \forall\ r>r^{\ast}\ ,\ \ \ \partial_{r}\alpha\underset
{r\rightarrow\infty}{\rightarrow}0\ . \label{ch2.5}%
\end{align}
The condition that $\alpha(r^{\ast})>\pi$ is necessary in order for $f^{2}$ to
be regular at $r^{\ast}$, as also the denominator in Eq. (\ref{inverscatt1})
has to vanish in order to compensate for the zero in the numerator:%
\[
\int_{0}^{r^{\ast}}x^{2}\left(  \partial_{x}\alpha\right)  \left(  \sin
\alpha(x)\right)  dx=0\ .
\]
In this case, it can be shown that for $r$ large enough $f(r)^{2}$ becomes
negative as well. Indeed, using the change of variable in Eqs. (\ref{varint1})
and (\ref{varint2}), one can evaluate the integral in the denominator in Eq.
(\ref{inverscatt1}) from $r^{\ast}$ up to infinity:%
\begin{align}
\int_{r^{\ast}}^{\infty}x^{2}\left(  \partial_{x}\alpha\right)  \left(
\sin\alpha(x)\right)  dx  &  =\int_{\alpha(r^{\ast})}^{0}\left(  x\left(
\alpha\right)  \right)  ^{2}\sin\alpha d\alpha\label{chh1}\\
&  =\int_{\alpha(r^{\ast})}^{\pi}\left(  x\left(  \alpha\right)  \right)
^{2}\sin\alpha d\alpha+\int_{\pi}^{0}\left(  x\left(  \alpha\right)  \right)
^{2}\sin\alpha d\alpha=I_{1}+I_{2}\ , \label{chh2}%
\end{align}
where%
\begin{align}
I_{1}  &  =\int_{\alpha(r^{\ast})}^{\pi}\left(  x\left(  \alpha\right)
\right)  ^{2}\sin\alpha d\alpha=-\int_{\pi}^{\alpha(r^{\ast})}\left(  x\left(
\alpha\right)  \right)  ^{2}\sin\alpha d\alpha>0\ ,\label{ineqint1}\\
I_{2}  &  =\int_{\pi}^{0}\left(  x\left(  \alpha\right)  \right)  ^{2}%
\sin\alpha d\alpha=-\int_{0}^{\pi}\left(  x\left(  \alpha\right)  \right)
^{2}\sin\alpha d\alpha<0\ . \label{ineqint2}%
\end{align}
Due to our hypothesis (see Eq. (\ref{ch2.5})) in the interval $\left]
r^{\ast},\infty\right[  $ the function $\alpha(r)$ is a decreasing function of
$r$ and, consequently, in the same interval the inverse function $x(\alpha)$
is a decreasing function of $\alpha$, so that the absolute value of the second
integral on the right hand side of Eq. (\ref{chh2}) (which is negative, see
Eq. (\ref{ineqint2})), is larger than the absolute value of the first integral
on the right hand side of Eq. (\ref{chh2}) (which is positive, see Eq.
(\ref{ineqint1})). Therefore, the integral on the left hand side of Eq.
(\ref{chh1}) is negative and this implies that also under the hypothesis in
Eqs. (\ref{ch1.2}) and (\ref{ch2.5}) $f^{2}$ becomes negative for $r$ large
enough and so solutions of the Gribov pendulum equations satisfying the strong
boundary conditions cannot be constructed. Following the same reasoning, it is
easy to show that also if one admits that $\alpha$ has more than one point
where the first derivative vanishes it is impossible to have $f^{2}$
everywhere positive from $0$ to $\infty$, smooth and well defined with, at the
same time, $\alpha$ fulfilling the strong boundary conditions. Indeed, nothing
would change by replacing the hypothesis in Eqs. (\ref{ch1.2}) and
(\ref{ch2.5}) with%
\begin{align*}
\alpha(0)  &  =2m\pi\ ,\ \alpha(r)\underset{r\rightarrow\infty}{\rightarrow
}2n\pi\ ,\ \left.  \partial_{r}\alpha\right\vert _{r=r_{i}}%
=0\ ,\ \ i=1,..,p\ ,\\
\ \partial_{r}\alpha &  >0\ \ \forall\ 0<r<r_{1}\ ,\ \partial_{r}%
\alpha<0\ \ \forall\ r_{1}<r<r_{2}\ ,...,\ \ \partial_{r}\alpha\underset
{r\rightarrow\infty}{\rightarrow}0:
\end{align*}
it is enough to repeat the previous argument starting with the last point in
which $\partial_{r}\alpha$ vanishes.

\subsection{Spacetime outside a black hole}

Let us now consider the cases of spherically symmetric spacetimes which
describe the exterior of a black hole. Since when one considers the Euclidean
version of a black hole spacetime, if there is a curvature singularity the
origin $r=0$ does not belong anymore to the spacetime itself, and the
condition in Eq. (\ref{origin}) (which ensures regularity at the origin in the
standard case) does not apply anymore.

In the case of black hole spacetimes Eq. (\ref{inverscatt1}) which expresses
the metric function $f$ in terms of the copy $\alpha$ changes as follows%
\begin{equation}
f(r)^{2}=\frac{\left(  r^{2}\partial_{r}\alpha\right)  ^{2}}{C+4\int_{r_{H}%
}^{r}x^{2}\left(  \partial_{x}\alpha\right)  \left(  \sin\alpha(x)\right)  dx}
\label{inverscatt2}%
\end{equation}
where $r_{H}$ is the radius of the event horizon. In the coordinate system
given in Eq. (\ref{Metrspher}), the event horizon can be characterized as a
pole of $f(r)^{2}$. Since we are considering regular copies, we have to
require that the derivative of $\alpha$ is bounded. Thus, one has to take
$C=0$ in Eq. (\ref{inverscatt2}) and, at the same time, the derivative of
$\alpha$ at $r_{H}$ does not vanish in such a way to get the desired pole. In
this case, it is easy to convince oneself that there is no choice of $\alpha$
such that $f(r)^{2}$\ is positive definite for $r>r_{H}$. To fix the ideas,
one can take a function $\alpha$ increasing monotonically from the value at
the horizon to $2\pi$ at infinity:
\begin{align}
\pi &  <\alpha(r_{H})<2\pi\ ,\label{cho1}\\
&  \ \alpha(r)\underset{r\rightarrow\infty}{\rightarrow}2\pi\ ,\ \partial
_{r}\alpha>0\ \ \forall\ r>r_{H}\ ,\ \ \ \partial_{r}\alpha\underset
{r\rightarrow\infty}{\rightarrow}0\ . \label{cho2}%
\end{align}
With this choice the integrand in the denominator in Eq. (\ref{inverscatt2})
does not change sign but one would have $f(r)^{2}<0$. If, instead, one assumes
that%
\begin{align}
0  &  \leq\alpha(r_{H})<\pi\ ,\label{cho1.5}\\
&  \ \alpha(r)\underset{r\rightarrow\infty}{\rightarrow}2\pi\ ,\ \partial
_{r}\alpha>0\ \ \forall\ r>r_{H}\ ,\ \ \ \partial_{r}\alpha\underset
{r\rightarrow\infty}{\rightarrow}0\ , \label{cho2.5}%
\end{align}
it can be shown that for $r$ large enough $f(r)^{2}$ becomes negative anyway.
Using the change of variable in Eqs. (\ref{varint1}) and (\ref{varint2}), the
integral in the denominator in Eq. (\ref{inverscatt2}) up to infinity reads%
\begin{align}
\int_{r_{H}}^{\infty}x^{2}\left(  \partial_{x}\alpha\right)  \left(
\sin\alpha(x)\right)  dx  &  =\int_{\alpha(r_{H})}^{2\pi}\left(  x\left(
\alpha\right)  \right)  ^{2}\sin\alpha d\alpha\label{int1}\\
&  =\int_{\alpha(r_{H})}^{\pi}\left(  x\left(  \alpha\right)  \right)
^{2}\sin\alpha d\alpha+\int_{\pi}^{2\pi}\left(  x\left(  \alpha\right)
\right)  ^{2}\sin\alpha d\alpha\ , \label{int2}%
\end{align}%
\[
\int_{\alpha(r_{H})}^{\pi}\left(  x\left(  \alpha\right)  \right)  ^{2}%
\sin\alpha d\alpha>0\ ,\ \ \int_{\pi}^{2\pi}\left(  x\left(  \alpha\right)
\right)  ^{2}\sin\alpha d\alpha<0\ .
\]
Since $\partial_{r}\alpha>0$, $x(\alpha)$ is an increasing function of
$\alpha$ so that the absolute value of the second integral on the right hand
side of Eq. (\ref{int2}) (which is negative) is larger than the first and
consequently the integral in Eq. (\ref{int1}) is negative. This implies that
for $r$ large enough $f(r)^{2}$ becomes negative. It is easy to see that the
same would happen with any choice in which $\alpha$ is a monotone function
$\forall\ r>r_{H}$ satisfying the strong boundary conditions. If one chooses a
function $\alpha$ which is not monotone, then $\partial_{r}\alpha$ would
vanish at least once for $r^{\ast}>r_{H}$. Thus, let us assume that
$\partial_{r}\alpha$ vanishes just once at $r^{\ast}$. Then, one has to
require that correspondingly also the denominator in Eq. (\ref{inverscatt2})
should vanish at $r^{\ast}$ in such a way to have a finite and positive
$f^{2}$. One can assume that $\alpha(r_{H})<\pi$ in such a way to ensure, at
least close to $r_{H}$, the positiveness of the integral:
\begin{align}
0  &  <\alpha(r_{H})<\pi\ ,\ \ \ \pi<\alpha(r^{\ast})<2\pi\ ,\label{chos1}\\
\partial_{r}\alpha &  >0\ \ \forall\ \ r_{H}<r<r^{\ast}\ ,\label{chos1.5}\\
\partial_{r}\alpha &  <0\ \ \forall\ r>r^{\ast}\ ,\ \ \alpha(r)\underset
{r\rightarrow\infty}{\rightarrow}0\ ,\ \ \ \partial_{r}\alpha\underset
{r\rightarrow\infty}{\rightarrow}0\ . \label{chos2}%
\end{align}
The above choice in Eqs. (\ref{chos1}) and (\ref{chos2}) for $\alpha(r^{\ast
})$ ensures that at least close to $r^{\ast}$ (the denominator of) $f^{2}$ is
positive. However, if one considers the integral from $r^{\ast}$ to infinity
in the denominator in Eq. (\ref{inverscatt2}) using the change of variable in
Eqs. (\ref{varint1}) and (\ref{varint2}):%
\begin{align}
\int_{r^{\ast}}^{\infty}x^{2}\left(  \partial_{x}\alpha\right)  \left(
\sin\alpha(x)\right)  dx  &  =\int_{\alpha(r^{\ast})}^{0}\left(  x\left(
\alpha\right)  \right)  ^{2}\sin\alpha d\alpha\label{int3}\\
&  =\int_{\alpha(r^{\ast})}^{\pi}\left(  x\left(  \alpha\right)  \right)
^{2}\sin\alpha d\alpha+\int_{\pi}^{0}\left(  x\left(  \alpha\right)  \right)
^{2}\sin\alpha d\alpha\ , \label{int4}%
\end{align}
where%
\begin{align*}
\int_{\alpha(r^{\ast})}^{\pi}\left(  x\left(  \alpha\right)  \right)  ^{2}%
\sin\alpha d\alpha &  =-\int_{\pi}^{\alpha(r^{\ast})}\left(  x\left(
\alpha\right)  \right)  ^{2}\sin\alpha d\alpha>0\ ,\\
\int_{\pi}^{0}\left(  x\left(  \alpha\right)  \right)  ^{2}\sin\alpha d\alpha
&  =-\int_{0}^{\pi}\left(  x\left(  \alpha\right)  \right)  ^{2}\sin\alpha
d\alpha<0\ ,
\end{align*}
once again one reaches the conclusion that the integral in Eq. (\ref{int3})
(and, consequently $f^{2}$) is negative because, for $r>r^{\ast}$, $x(\alpha)$
is a decreasing function of $\alpha$ so that the absolute value of the second
integral on the right hand side of Eq. (\ref{int4}) is larger than the
absolute value of the first integral. It is easy to see that the same
conclusion would hold in the case in which the derivative of $\alpha$ would
vanish at more than one point. Hence, also if one assumes that $\alpha$ is not
monotone, for $r$ large enough $f^{2}$ becomes negative. Hence, on spherically
symmetric black hole spacetimes as in Eq. (\ref{Metrspher}) solutions of the
Gribov pendulum equation representing copies of the vacuum satisfying the
strong boundary conditions cannot appear.

\subsection{ The space-time outside a star}

In the previous subsections it has been shown that both, on spherically
symmetric regular spacetimes and on spherically symmetric black hole
spacetimes one cannot construct solutions satisfying the strong boundary
conditions. The main technical reason is that both in Eq. (\ref{inverscatt1})
and in Eq. (\ref{inverscatt2}) one has to take $C=0$. In the first case, this
is necessary in order to achieve a spacetime which is regular at the origin,
while in the second case $C=0$ ensures the appearance of the black hole
horizon at $r_{H}$. Indeed, when $C=0$ all the previous arguments on the
change of sign of the integral in the denominator of the expression for
$f^{2}$ work.\textbf{ }However, the situation is radically different in the
cases in which the background metric in Eq. (\ref{Metrspher}) represents, for
instance, the exterior of a spherically symmetric star. In the case of a
spacetime representing the exterior of a star Eq. (\ref{inverscatt1}) which
expresses the metric function $f$ in terms of the copy $\alpha$ changes as
follows%
\begin{equation}
f(r)^{2}=\frac{\left(  r^{2}\partial_{r}\alpha\right)  ^{2}}{C+4\int_{r_{S}%
}^{r}x^{2}\left(  \partial_{x}\alpha\right)  \left(  \sin\alpha(x)\right)
dx}\ , \label{inverscatt3}%
\end{equation}
where $r_{S}$ is the coordinate radius of the star. Unlike the black hole case
in which one has to require that $f^{2}$ in Eq. (\ref{Metrspher}) has a pole
at $r_{H}$, in the case of a spacetime representing the exterior of a star one
has to require that $f^{2}$ evaluated at $r_{S}$ should be finite and
non-vanishing. This fact has the highly non-trivial consequence that, in this
case, $C$ can be chosen to be non-vanishing and this allows one to construct
infinite examples of curved backgrounds supporting copies of the vacuum
satisfying the strong boundary conditions. To see this, one can consider, for
instance, a monotone function $\alpha$ varying from the value at the horizon
to $2\pi$ at infinity:
\begin{align}
\pi &  <\alpha(r_{S})<2\pi\ ,\ \ \label{star1}\\
&  \ \alpha(r)\underset{r\rightarrow\infty}{\rightarrow}2\pi\ ,\ \partial
_{r}\alpha>0\ \ \forall\ r>r_{S}\ ,\ \ \ \partial_{r}\alpha\underset
{r\rightarrow\infty}{\rightarrow}0\ . \label{star2}%
\end{align}
Since in the denominator of Eq. (\ref{inverscatt3}) $C$ can be chosen at will,
one can take a positive value of $C$ large enough to prevent any change of
sign in the denominator:%
\[
C>4\int_{r_{S}}^{\infty}\left\vert x^{2}\left(  \partial_{x}\alpha\right)
\left(  \sin\alpha(x)\right)  dx\right\vert \ ,
\]
the above condition also implies that the integral in the denominator of Eq.
(\ref{inverscatt3}) has to converge. If the above constraint is satisfied,
then $f^{2}$ is everywhere positive and the corresponding background metric
supports, by construction, a copy of the vacuum satisfying the strong boundary
conditions. This argument shows that there is a huge freedom in constructing
background supporting such copies of the vacuum since the function $\alpha$,
besides the conditions in Eqs. (\ref{star1}) and (\ref{star2}), can be chosen arbitrarily.

A consequence of the present analysis is that a space-time supporting copies
of the vacuum of the form in Eq. (\ref{gaugtransf}) with strong boundary
condition, can not be asymptotically Minkowski (as it can be verified directly
by expanding, for large $r$, Eq. (\ref{inverscatt3})) whereas it can be
asymptotically AdS provided%
\[
\alpha(r)\underset{r\rightarrow\infty}{\rightarrow}2\pi+\frac{k}{r^{2}%
}+O(1/r^{3})\ ,\ \partial_{r}\alpha\underset{r\rightarrow\infty}{\rightarrow
}-\frac{2k}{r^{3}}+O(1/r^{4})\ ,
\]
$k$ being a real constant.\newline It is worth pointing out that in this
construction the value of $\alpha$ at $r_{S}$ must be different from its value
at infinity, otherwise its derivative would be somewhere zero and the metric
would be singular there. This implies that these vacuum Gribov copies cannot
have trivial winding \cite{SS05}. Of course, as it has been already
emphasized, even when the copy has a non-trivial winding the corresponding
gauge transformation represents a proper gauge transformation which cannot be
discarded provided the strong boundary conditions are satisfied as it does not
change the value of the observables.

Moreover, in the cases in which it is possible to find a copy also for $0\leq
r<r_{S}$ (the region which may represent the interior of the star), then one
could match in a smooth ($C^{1}$ ) way the interior and the exterior copies to
get a globally defined copy without any winding\footnote{It is worth
remembering that in the expression for the winding number of a gauge
transofrmation $U$ only first derivetives of $U$ appear so that the winding
number is well defined whenever $U$ is $C^{1}(M)$ ($M$ being the spacetime of
interest).}. The $C^{1}$ matching of the copy appears to be possible due to
the freedom given by the integration constants $A$ and $B$ appearing in
(\ref{fixedpoint}).

\section{Conclusions and further comments}

In this paper we analyzed the curved generalization of the Gribov pendulum in
the Coulomb gauge on static spherically symmetric space-times. Using tools of
non-linear functional analysis, we explored the issue of existence and
uniqueness of solution of the Gribov pendulum on asymptotically AdS spacetimes
in terms of an effective Schr\"{o}dinger equation. Furthermore, we constructed
necessary conditions in order for a curved static spherically symmetric
background to admit copies of the vacuum satisfying the strong boundary
conditions. An interesting consequence of the present analysis is that
asymptotically Minkowski spacetimes do not admit vacuum copies of the Gribov
form in Eq. (\ref{gaugtransf}) fulfilling the strong boundary conditions. This
strongly suggests that as it happens in flat spacetime, QCD at perturbative
level is not affected by Gribov ambiguities in such cases.

\bigskip

The situation changes dramatically when one considers asymptotically AdS
spacetimes. In these cases vacuum copies can appear depending on the structure
of the interior bulk spacetime. In particular black holes do not admit vacuum
copies of the Gribov form (\ref{gaugtransf}) whereas an asymptotically AdS
spacetime containing, for instance, a star does admit vacuum copies satisfying
the strong boundary conditions. An interesting issue arises if one considers
the gravitational collapse of a star to a black hole with AdS asymptotics.
This would imply a sudden change in the size of the Gribov horizon. According
to the Gribov-Zwanziger approach this would imply a sudden change in the
infrared behavior of QCD in these spacetimes.

Assuming the validity of the Gribov-Zwanziger procedure (which is supported by
lattice data), the strong dependence of QCD on the structure of the interior
bulk spacetime may have also interesting consequences for the AdS/CFT
correspondence. Indeed, the Gribov problem affects directly the gluon
propagator and the AdS/CFT correspondence is a statement about gauge invariant
operators. Notwithstanding, in the present framework the presence of Gribov
copies is also relevant as far as confinement is concerned, which implies that
one should also expect that gauge invariant operators will be modified.

Our results suggest that the Gribov-Zwanziger confinement picture is stable
under perturbations of the flat background metric that do not change the
asymptotic structure. The reason is that for the family of copies considered
here, the asymptotically flat case behaves in a very similar manner as the
flat case. Our results also suggest that the pattern of appearance of Gribov
copies inside a star (as well as outside a star with AdS asymptotic) should be
very different from the flat case due to the presence of an intrinsic length
scale of the problem (the radius of the star). In particular, in a curved
background in which also the vacuum possesses copies satisfying the strong
boundary conditions the very notion of asymptotic freedom could change
dramatically. In fact, the usual scenario on flat spaces is that the deep
ultra-violet region corresponds to the trivial vacuum $A_{\mu}=0$, which is
free of Gribov copies satisfying the strong boundary conditions. Thus, the
absence of a region of the functional space of the gauge potential $A_{\mu}$
free of strong copies could be interpreted as the absence of a
perturbative-deconfined region. This problem could be of great interest also
because of its astrophysical implications and it is currently under
investigation. 

It is also worth pointing a further interesting possibility related with the
present scenario. Our results suggest that when a star living in a
asymptotically AdS space-time undergoes a gravitational collapse to a black
hole, the strong Gribov copies of the vacuum may disappear in very much the
same way as it happens in the flat case. This suggests that a gravitational
collapse in an asymptotically AdS space-time could induce a sort of phase
transition for the QCD degrees of freedom outside the star collapsing to the
black hole corresponding to the appearance, in the black hole phase, of a
Gribov horizon around the trivial vacuum $A_{\mu}=0$. This issue and also how
such a phase transition may be visible in the dual boundary theory is not at
all clear up to now and will be issue of further investigation.

\bigskip

\section{Acknowledgments}

We thank Andr\'{e}s Anabal\'{o}n for useful comments. This work is partially
supported by FONDECYT grants 11080056, 11090281, and 1110167, and by the
\textquotedblleft Southern Theoretical Physics Laboratory\textquotedblright%
\ ACT-91 grant from CONICYT. The Centro de Estudios Cient\'{\i}ficos (CECs) is
funded by the Chilean Government through the Centers of Excellence Base
Financing Program of CONICYT. F. C. is also supported by Proyecto de
Inserci\'{o}n CONICYT 79090034, and by the Agenzia Spaziale Italiana (ASI).


\begin{thebibliography}{99}                                                                                               %


\bibitem {Benguria:1976in}R.~Benguria, P.~Cordero and C.~Teitelboim,
Nucl.\ Phys.\ B \textbf{122} (1977) 61.


\bibitem {wittenjones}E. Witten, \textit{Comm. Math. Phys}. \textbf{121},
(1989) 351.

\bibitem {HRT}A. Hanson, T. Regge, C. Teitelboim, \textquotedblright%
\textit{Constrained Hamiltonian Systems}\textquotedblright, Accademia
Nazionale dei Lincei, Roma (1976).

\bibitem {DeW03}B. S. DeWitt, "\textit{Global approach to quantum field
theory}" Vol. 1 and 2, Oxford University Press (2003).

\bibitem {Gri78}V.N. Gribov, \textit{Nucl. Phys}. \textbf{B 139} (1978) 1.

\bibitem {singer}I. M. Singer, \textit{Comm. Math. Phys}. \textbf{60} (1978), 7.

\bibitem {Zw82}D. Zwanziger, Nucl. Phys. \textbf{B 209}, (1982) 336.

\bibitem {Zw89}D. Zwanziger, Nucl. Phys. \textbf{B 323}, (1989) 513.

\bibitem {DZ89}G. F. Dell'Antonio, D. Zwanziger, Nucl. Phys. \textbf{B 326},
(1989) 333.

\bibitem {Zwa96}D. Zwanziger, \textit{Nucl. Phys.} \textbf{B 518} (1998) 237;
\textit{Phys. Rev. Lett}. \textbf{90} (2003) 102001.

\bibitem {Va92}P. van Baal, Nucl. Phys. \textbf{B 369}, (1992) 259.

\bibitem {SS05}R. F. Sobreiro, S. P. Sorella, "\textit{Introduction to the
Gribov Ambiguities In Euclidean Yang-Mills Theories}" arXiv:hep-th/0504095; D.
Dudal, M. A. L. Capri, J. A. Gracey, V. E. R. Lemes, R. F. Sobreiro, S. P.
Sorella, R. Thibes, H. Verschelde, "\textit{Gribov ambiguities in the maximal
Abelian gauge}" arXiv:hep-th/0609160.

\bibitem {EPZ04}G. Esposito, D.N. Pelliccia, F. Zaccaria, Int. J. Geom. Meth.
Mod. Phys. 1 (2004) 423.

\bibitem {Fuj}K. Fujikawa, \textit{Nucl. Phys}.\textbf{ B223} (1983) 218.

\bibitem {Sor1}D.Dudal, S.P.Sorella, N.Vandersickel, H.Verschelde,
\textit{Phys.Rev}\textbf{.D77 }(2008), 071501.

\bibitem {Sor2}D. Dudal, J. Gracey, S. P. Sorella, N. Vandersickel, H.
Verschelde, \textit{Phys.Rev}\textbf{.D78 }(2008), 065047.

\bibitem {BaSo}L. Baulieu, S. P. Sorella, \textit{Phys. Lett}.\textbf{ B 671}
(2009) 481.

\bibitem {CGO}F.~Canfora, A.~Giacomini and J.~Oliva,
Phys.\ Rev.\ D \textbf{82}, 045014 (2010) [arXiv:1004.2718 [hep-th]].


\bibitem {ACGO}A. Anabalon, F.~Canfora, A.~Giacomini and J.~Oliva,
Phys.\ Rev.\ D \textbf{83}, 064023 (2011).

\bibitem {QUARKSTAR}A. R Zhitnitsky, JCAP 2003 (10) 010; P.Jaikumar, S. Reddy,
A. Steiner, \textit{Phys. Rev. Lett.} \textbf{96}, 041101 (2006); J. E.
Truemper, V. Burwitz, F. Haberl, V. E. Zavlin, \textit{Nucl. Phys.} \textbf{B}
(Proceedings Supplements) 132: 560 (2004).

\bibitem {maldacena}J. Maldacena, \textit{Adv.Theor.Math.Phys.}2:231-252,1998;
\textit{Int.J.Theor.Phys.}38:1113-1133,1999

\bibitem {magoo}O.~Aharony, S.~S.~Gubser, J.~M.~Maldacena, H.~Ooguri and
Y.~Oz,
Phys.\ Rept.\ \textbf{323}, 183 (2000) [arXiv:hep-th/9905111].


\bibitem {adscondmat}J. H. Schwarz, "Some Recent Progress in AdS/CFT"
arXiv:1006.1670; S. Sachdev, "Condensed matter and AdS/CFT" arXiv:1002.2947.

\bibitem {berger}M. Berger, \textit{nonlinearity and functional analysis},
Academic press, 1977.

\bibitem {giltrud}D. Gilbarg, N. S. Trudinger, Elliptic partial differential
equations of second order, Springer-Verlag, 1983.

\bibitem {heniey}F. S. Henyey, Phys. Rev. D \textbf{20}, 1460 (1979).
\end{thebibliography}
\end{document}